\documentclass[aps,twocolumn,showpacs,amsmath,amssymb]{revtex4}
\usepackage[T2A]{fontenc}
\usepackage{amsmath}
\usepackage{amssymb}
\usepackage{graphicx}
\usepackage{euscript}

\newlength{\pictwidth}
\newcommand{\aver}[1]{{\left< #1 \right>}}

\begin{document}
\title{Sensitivity of ray dynamics in an underwater sound channel
to vertical scale of longitudinal sound-speed variations}

\author{D.V. Makarov}
\email{makarov@poi.dvo.ru}
\author{M.Yu. Uleysky}
\author{M.Yu. Martynov}
\affiliation{V.I.Il'ichev Pacific Oceanological Institute \\
of the Russian 
Academy of Sciences, 690041 Vladivostok, Russia}

\begin{abstract}
We investigate 
sound ray propagation in a range-dependent underwater
acoustic waveguide.
Our attention is focused on sensitivity of ray dynamics
to the vertical structure of a sound-speed perturbation
induced by ocean internal waves.
Two models of longitudinal sound-speed variations are considered: 
a periodic inhomogeneity and a stochastic one.
It is found that vertical oscillations of a sound-speed perturbation can affect rays in a resonant manner.
Such resonances give rise to chaos in certain regions of phase space.
It is shown that stability of steep rays, being observed in experiments,
is connected with suppression of resonances
in the case of small-scale vertical sound-speed oscillations.

\end{abstract}
\maketitle

\section{Introduction}

In recent two decades 
exponential divergence of ray trajectories with
an infinitesimal difference in initial conditions,
namely ray chaos \cite{Chig,Kras,Pal,Ufn,Review},
stands as a topical problem in underwater acoustics.
Practical sense of studying ray dynamics under chaotic conditions
is mainly dealed with the hydroacoustical tomography --- monitoring of the
ocean using sound signals \cite{MunkW, Spies1, Spies2, Akul, IEEE, Akust}.
Ray chaos raises difficulties in extracting 
an information about environment
from data of long-range acoustical experiments
using the traditional schemes of the tomography.
Extreme sensitivity of chaotic rays to initial conditions
sets the ``predictability horizon'' on ray motion.
Furthermore, ray chaos leads
to multiplying of eigenrays and causes breakdown
of the ray perturbation theory at long ranges \cite{Smith1, Smith2, Tap}.
Nowadays development of efficient tomographical methods
seeks the detailed description of ray chaos and its manifestations
in wavefield structure with the goal to obviate
posed restrictions, if it is possible.
The helpful feature in this way is surprising stability of the early portion
of a received pulse.
This phenomenon was observed in a number of experiments realized by NPAL group (see \cite{Spind}
and references therein), 
for instance, in the well-known AET experiment \cite{Worc}.
The numerical simulation of ray propagation in the AET
environment shows that characteristic Lyapunov exponents of steep rays
are less significantly than those of near-axial ones \cite{AET}. 
Such a picture indicates suppressing of 
diffusion in the range of large values of the action variable
and strong chaos in the range of small ones.
However, if the internal wave field is modeled as a sum of horizontal modes only,
the range small values of the action corresponds to stable motion \cite{Chaos}.
The primary goal of the present work is to examine
a role of vertical sound-speed variations in ray properties.
The most attention is paid to the model of a waveguide
with a deterministic periodic perturbation of a sound-speed profile.
Although this model is idealistic, it yields a simple description 
of chaotic dynamics in the framework of the KAM (Kolmogorov--Arnold--Moser)
theory.
The case of the stochastic perturbation is also considered
to make our research more complete.

The paper is organized as follows.
In Section II we describe briefly the model of a sound-speed profile,
which we used in numerical experiments. Section III is devoted to analysis
of ray dynamics with a periodic spatial inhomogeneity of a waveguide. 
The sensitivity of ray propagation to the vertical scale of a perturbation
is examined by computing Poincar\'e sections and timefronts of a received
pulse.
Section IV is concerned with ray propagation
in a stochastically inhomogeneous waveguide. In Section V we summarize
and discuss our results. The Appendix contains the exact expressions
for the action--angle variables which was introduced
for the model of an unperturbed waveguide.

\section{Model of a waveguide}

Consider a two-dimensional acoustic waveguide
in the deep ocean
with the sound speed $c$ being a smooth function of the depth $z$ and the range $r$.
One-way sound ray trajectories satisfy the canonical Hamilton 
equations
\begin{equation}
\dfrac{dz}{dr}=\dfrac{\partial H}{\partial p},\quad
\dfrac{dp}{dr}=-\frac{\partial H}{\partial z},
\label{sys}
\end{equation}
with the Hamiltonian 
\begin{equation}
H=-\sqrt{n^2(z,\,r)-p^2},
\label{ham-init}
\end{equation}
where $n=c_0/c(z,r)$ is the refractive index, 
$c_0$ is some reference sound speed,
$p=n\sin\phi$ is so-called ray momentum, and $\phi$ is ray grazing angle.
As sound speed changes slightly with depth, i.~e.
$|c(z,r)-c_0|/c_0\ll 1$, one
permits to rewrite the Hamiltonian (\ref{ham-init}) in the paraxial
(small--angle) approximation
\begin{equation}
H=-1+\dfrac{p^2}{2}+\dfrac{\Delta c(z)}{c_0}+
\dfrac{\delta c(z,\,r)}{c_0},
\label{ham01}
\end{equation}
where $\Delta c(z)=c(z)-c_0$, $\delta c(z,\,r)$ is
a sound-speed perturbation along a waveguide,
$\delta c_\text{max}\ll\Delta c_\text{max}$.
It is clearly seen from (\ref{ham01}) that ray propagation
in a range-dependent underwater sound channel is equivalent
to motion of a nonlinear oscillator driven by an external force.
According to this analogy, a sound-speed profile plays role
of a potential well, $p$ is the analog to mechanical momentum, and
$r$ is the timelike variable.

In Ref. \cite{Chaos,Dan,PhD} we have introduced the realistic model of a background sound-speed 
profile
\begin{equation}
c(z)=
c_0\left[1-\dfrac{b^2}{2}(\mu-e^{-az})(e^{-az}-\gamma)\right],\quad
0\le z\le h,
\label{prof}
\end{equation}
where $\gamma =\exp(-ah)$, $\mu=1.078$,
$a=0.5$~km$^{-1}$, $b=0.557$,
$h=4.0$~km is the depth of the ocean bottom, $c_0=c(h)=1535$~m/s.

\begin{figure}[!ht]
\begin{center}
\centerline{
\includegraphics[width=0.15\textwidth]{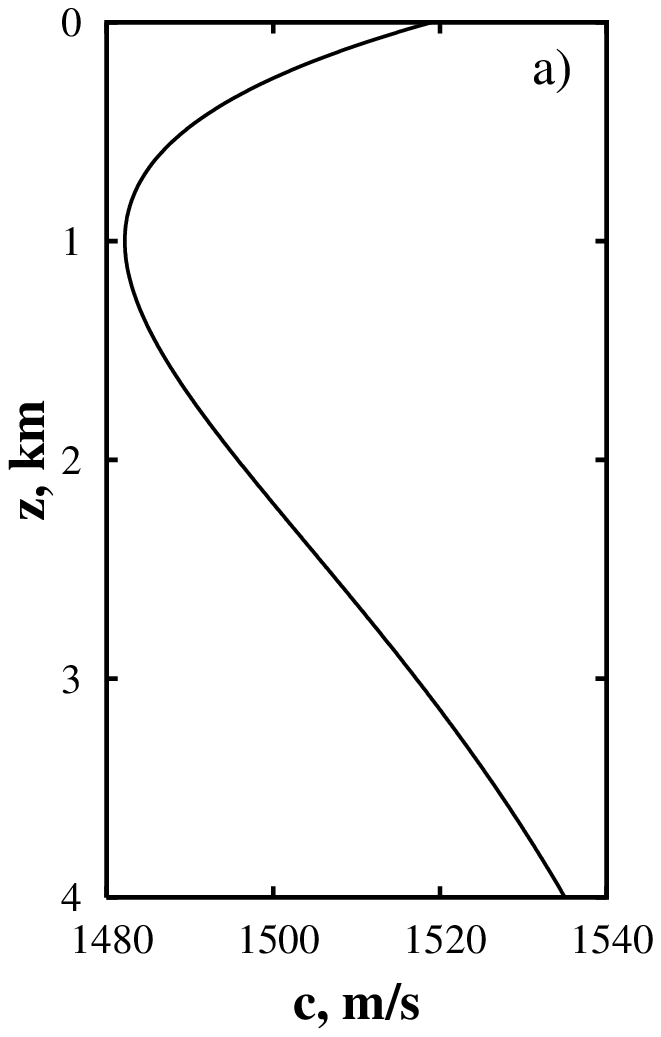}
\includegraphics[width=0.15\textwidth]{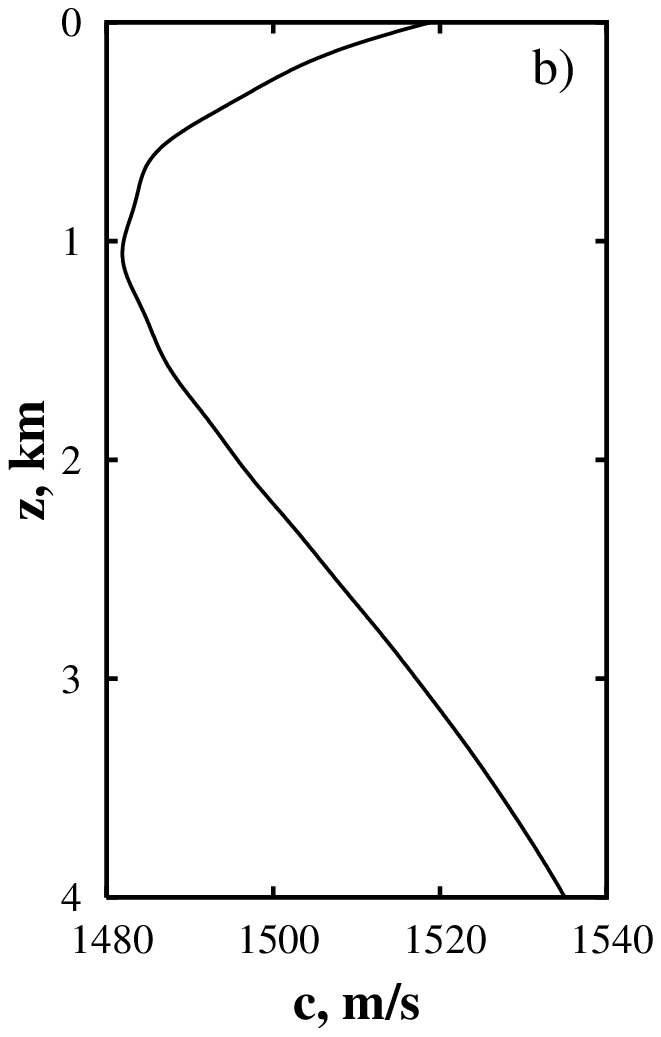}
\includegraphics[width=0.15\textwidth]{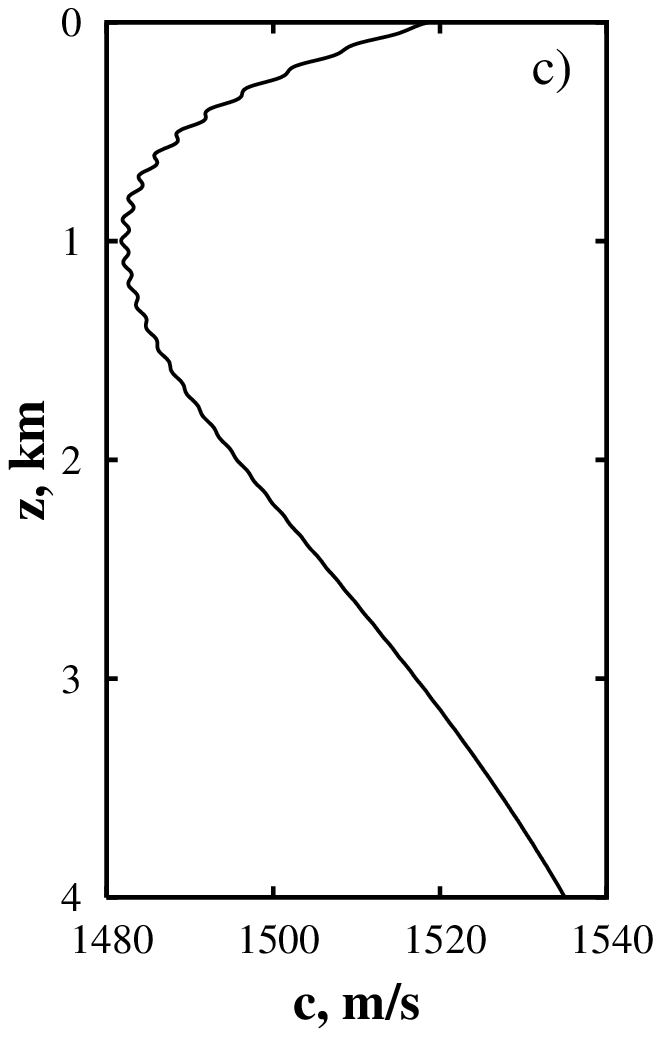}}
\caption{Sound-speed profiles at the range $R=0$:
a) without perturbation,
b) $k_z=2\pi/1.0$~km$^{-1}$,
c) $k_z=2\pi/0.2$~km$^{-1}$.
}
\label{profs}
\end{center}
\end{figure}

For convenience we introduce the quantity $E=1+H(\varepsilon=0)$
which we assign as the ``energy'' of ray oscillations in a waveguide.
Rays, propagating in the waveguide (\ref{prof})
over long distances, can be partitioned
into two classes according to the value of $E$.
The first class corresponds to those rays which
don't reach neither the surface nor the bottom.
For these rays $E<E_r$, where $E_r$ is given by
\begin{equation}
E_{r}=\frac{b^2}{2}(1-\mu)(1-\gamma).
\label{Er}
\end{equation}
Rays, propagating
with reflections from the surface
but without an interaction with the lossy bottom, relate to the second class.
In this case energy satisfies to the inequality $E_r\le E<0$.
Rays with $E\ge 0$ reach the ocean bottom.
Just as we consider the long-range sound propagation,
number of the reflections from the bottom is large enough and, therefore,
these rays cannot survive in the waveguide over long
distances and should be eliminated.
Consequently, we can define the phase trajectory of the ray
with $E=0$ as the ``separatrix'' of unperturbed motion.
As energy $E$ isn't an invariant along the
trajectory in a range--dependent waveguide,
rays can transfer from one class to another.
Whenever $E$ becomes larger than zero,
the ray reaches the bottom and escapes from the waveguide. Ray escaping
is closely connected to transport properties of phase space \cite{Meiss,
PhysD}.
In certain cases
the portion of escaping rays
enables to evaluate the rate of diffusion in phase space \cite{PhD}.
In addition to that, we note that
ray escaping takes place in waveguides having
various physical nature, for instance,
a similar effect is observed for the lasing modes in the
chaotic dielectric microcavities \cite{Lee}. 

For the unperturbed problem we are able to to reduce
the ray equations to a more convenient form
in terms of canonical action--angle variables \cite{Ufn,Review,LanLif}.
The respective transformation is the following
\begin{equation}
I=\frac{1}{2\pi}\oint p\,dz,\quad
\vartheta=\frac{\partial}{\partial I}\int\limits_{z_0}^z p\,dz.
\label{acangle}
\end{equation}
The exact analytical expressions for the action--angle variables
and for the inverse transition to the variables ($p$--$z$) are presented
in the Appendix. The action variable $I$ measures the steepness of the ray
and is equal to 0 for the purely horizontal ray. 
Besides that, the action determines the mode the ray corresponds to
by means the Bohr-Zommerfeld quantization rules 
\begin{equation}
k_0I_m=\left\{
\begin{aligned}
m+\frac{1}{2},\quad 0\le E<E_r,\\
m-\frac{1}{4},\quad E_r\le E<0,
\end{aligned}
\right.
\label{BZ}
\end{equation}
where $k_0=2\pi\Omega/c_0$ is wavenumber of the propagating wave, $\Omega$ is the career frequency.
The angle variable $\vartheta$ serves as a phase of ray oscillations 
in a waveguide.

\begin{figure}[!htb]
\centerline{\includegraphics[width=0.4\textwidth,clip]{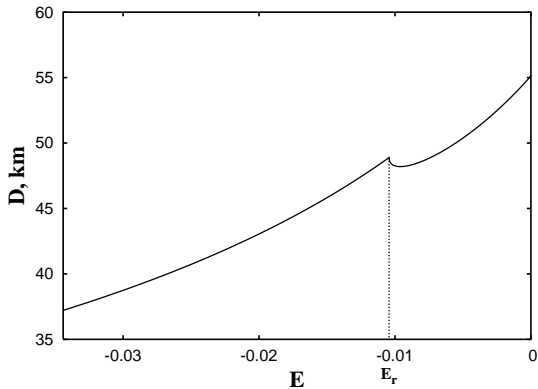}}
\caption{The cycle length of the ray path $D$ versus the energy $E$
of ray oscillations in the unperturbed waveguide.}
\label{d-e}
\end{figure}

The ray equations in terms of the action-angle variables are written
as follows
\begin{equation}
\begin{aligned}
\frac{dI}{dr}=&-\frac{1}{c_0}\frac{\partial(\delta c)}{\partial\vartheta},\\
\frac{d\vartheta}{dr}=&\frac{2\pi}{D}+
\frac{1}{c_0}\frac{\partial(\delta c)}{\partial I},
\end{aligned}
\label{dot-i} 
\end{equation}
where the sound-speed perturbation $\delta c$ is expressed as a function
of $I$ and $\vartheta$, and $D$ is 
the ray cycle length (ray double loop).
The ray cycle length in the waveguide (\ref{prof})
is given by the formulas
\begin{equation}
D=\left\{
\begin{aligned}
&\frac{2\pi}{ab\sqrt{\mu\gamma-2E/b^2}},\quad E\le E_r\\
&\dfrac{\left(\pi+
2\arcsin\dfrac{\mu+\gamma-2\mu\gamma+4E/b^2}{Q}\right)}
{ab\sqrt{\mu\gamma-2E/b^2}},\quad E>E_r.
\end{aligned}
\label{d1}
\right.
\end{equation}
As it is shown from Fig.~\ref{d-e},
$D$ increases with increasing $E$
from 37~km to 55~km \cite{Chaos}. Since the derivative $dD/dE$ has
two isolated zeroes near $E=E_r$, the system of
nonlinear coupled equations (\ref{sys}) is locally degenerate. 
Such systems are known to exhibit chaotic behavior even under
infinitesimal perturbations \cite{Review}.

In accordance with Fermat's principle, ray arrival time to
a point $R$ along a waveguide is calculated with the help of
the Lagrangian $L$
\begin{equation}
t=\dfrac{1}{c_0}\int\limits_0^RL\, dr
=\dfrac{1}{c_0}\int\limits_0^R (p^2-H)\,dr.
\label{time}
\end{equation}
At sufficiently long ranges, $R/D\gg 1$, the Lagrangian $L$ may be
considered as a function of the action
\begin{equation}
L(I) = 2\pi \dfrac{I}{D(I)}-H_0(I).
\label{lag_i}
\end{equation}
Following to (\ref{time}), ray arrival time to the point $R$
along a range-dependent waveguide is given by
\begin{equation}
t=\dfrac{R}{c_0}\aver{L(I)},
\label{tcom}
\end{equation}
where $\aver{\dots}$ means averaging over $r$.

\section{Periodic sound-speed perturbation}

\subsection{Theoretical analysis}

In this section we consider the periodic sound-speed perturbation given in the form
\begin{equation}
\frac{\delta c(z,\,r)}{c_0}=\varepsilon V(z)\cos{(k_rr+k_zz)},
\label{dc}
\end{equation}
where $\varepsilon\ll 1$ and $V(z)$ is a smooth function playing role
of an ``envelope'' of sound-speed variations.
In numerical experiments we use the following expression for $V(z)$ 
\begin{equation}
V(z)=\frac{z}{B}\,e^{-2z/B},
\label{fz}
\end{equation}
where $B=1$~km.
The sound-speed perturbation (\ref{dc}) 
in terms of the action-angle variables reads
\begin{equation}
\frac{\delta c(z,\,r)}{c_0}=\varepsilon V(I,\,\vartheta)
\cos{\Phi}
\label{dcI}
\end{equation}
where the phase $\Phi=k_rr+k_zz$ varies 
along the ray path with the angle-dependent spatial frequency
\begin{equation}
\frac{d\Phi}{dr}=k(I,\,\vartheta)=k_r+k_zp(I,\,\vartheta).
\label{k1}
\end{equation}
%
%
%
The function $V(I,\,\vartheta)$ can be decomposed
into the Fourier series over the cyclic variable $\vartheta$
\begin{equation}
V(I,\,\vartheta)=\sum_{m=1}^\infty
V_{m}(I)\,\cos{m\vartheta}.
\label{v_ryad}
\end{equation}
Then the ray equations (\ref{dot-i}) take form
\begin{equation}
\begin{aligned}
\dfrac{dI}{dr}=&\frac{\varepsilon}{2}\Biggl(\Biggr.
\sum_{m}mV_{m}(\sin{\Psi_m^-}+\sin{\Psi_m^+})+ \\
&+\frac{k_zp}{\omega}\sum_{m}V_{m}(\sin{\Psi_m^+}-\sin{\Psi_m^-})
\Biggl.\Biggr),
\end{aligned}
\label{i}
\end{equation}
\begin{equation}
\begin{aligned}
\dfrac{d\vartheta}{dr}=&\omega+\frac{\varepsilon}{2}\biggl(\biggr.
\sum_{m}\frac{V_{m}}{dI}(\cos{\Psi_m^-}+\cos{\Psi_m^+})- \\
&-k_z\sum_{m}V_{m}\frac{\partial z}{\partial I}(\sin{\Psi_m^-}+\sin{\Psi_m^+})
\biggl.\biggr),
\end{aligned}
\label{th}
\end{equation}
where $\omega=2\pi/D$ and
$\Psi_m^{\pm}=m\vartheta\pm k_rr\pm k_zz(I,\,\vartheta)$.
The stationary phase conditions $d\Psi_m^{\pm}/dr\simeq 0$
imply the first--order nonlinear resonances
\begin{equation}
m\omega(I)-k_r-k_zp(I,\,\vartheta)\simeq 0,
\label{res1}
\end{equation}
\begin{equation}
m\omega(I)+k_r+k_zp(I,\,\vartheta)\simeq 0.
\label{res2}
\end{equation}
%

We intent to investigate how interrelation between different length scales
of a perturbation influences the rays.
Three qualitatively different regimes of ray motion
can be separated according to the ratio of $k_z$ and $k_r$:

1) $|k_zp_\text{max}|\ll k_r$;

2) $|k_zp_\text{max}|\simeq k_r$;

3) $|k_zp_\text{max}|\gg k_r$, \\
where $p_\text{max}$ is the ray momentum at the channel axis.

The first regime corresponds to a horizontal sound-speed perturbation
(see, for instance \cite{Pal, Ufn, Chaos, Smir}).
In this case the resonant condition is the simplest 
\begin{equation}
m\omega(I=I_\text{res})=lk_r,
\label{rescond}
\end{equation}
where $l$ and $m$ are integers. 
The action of a ray under a resonance (\ref{rescond}) oscillates with range.
These oscillations are described by the universal Hamiltonian
of nonlinear resonance \cite{Ufn, Chirikov}
\begin{equation}
H_u=m\left(\frac{1}{2}\,\bigl|\omega'_I(I_\text{res})\bigr|\left(\Delta I\right)^2
+\varepsilon|V_m|\cos{\Psi_m^-}\right),
\label{universal}
\end{equation}
where $\omega'_I(I_\text{res})=d\omega(I_\text{res})/dI$,
$\Delta I=I-I_\text{res}$. The width of an isolated
resonance in terms of the spatial frequency of a ray trajectory
is approximately estimated as
\begin{equation}                    
\Delta\omega_\text{max}=|\omega'_I|\Delta I_\text{max}=
2\sqrt{\varepsilon|\omega'_I|V_{m}},
\label{shir}
\end{equation}
where $\Delta I_\text{max}$ is the width of the resonance in terms of 
the action variable. 
In accordance with (\ref{tcom}),
if the distance between the source and the receiver is large compared with
the period of action oscillations, all the rays, being captured in a given resonance,
have close arrival times.
Consequently, sharp peaks and gaps
arise in distribution of ray arrival times at a receiver \cite{Chaos,Smir,PJTF}.

In accordance with Chirikov's criterion \cite{Chirikov}, 
global chaos arises if 
\begin{equation} 
\frac{\Delta\omega_\text{max}}{\delta\omega}\simeq 1,
\label{Chi} 
\end{equation} 
i.~e., if two nonlinear resonances, centered at $\omega$ and 
$\omega + \delta \omega$, overlap. 
Those resonances that overlap slightly
form islands in phase space, areas of stable ray motion in the chaotic sea.
The distance between the 
resonances of the $m$-th and $m+1$-th orders in terms of spatial frequency is 
equal to 
\begin{equation}
\delta\omega=\frac{k}{m}-\frac{k}{m+1}\simeq\frac{\omega^2}{k}
\propto D^{-2},
\label{rasst}
\end{equation}
and decreases rapidly with increasing $D$. Hence chaotic properties
of rays under a horizontal sound-speed perturbation depend
on the shape of the unperturbed profile  
defining the function $D(E)$ \cite{BV}.

Under the second regime the condition (\ref{res2}) can be satisfied,
that corresponds to capturing a ray into the resonance with 
vertical oscillations of a sound-speed perturbation.
This phenomenon is similar to the 
``wave--particle'' resonance \cite{LichLib, Zas}. Hereafter we will call 
this sort of resonances as ``vertical resonance''.
Vertical resonance arises
at some fixed point where the ray momentum is equal to 
\begin{equation}
p_\text{res}(I)=-\frac{m\omega(I)+k_r}{k_z}.
\label{res20}
\end{equation}
%


Define the $\mu$-vicinity of a given ray as 
a small region along a ray trajectory, where the detuning of
vertical resonance doesn't exceed some small value $\mu$, i. e.
\begin{equation}
|\Delta|=|m\omega+k|=k_z|p-p_\text{res}|\le\mu.
\label{mu}
\end{equation}
Influence of vertical resonance depends on how long the ray
passes through the $\mu$-vicinity,
Consequently, the maximal influence is expected, when 
the $\mu$-vicinity includes the channel axis where $dp/dr=0$ and
$p=\text{max}$.
Therefore, vertical resonance
affects mainly those rays which obey to the equation
\begin{equation}
p_\text{max}(I)=-\frac{m\omega(I)+k_r}{k_z}.
\label{res21}
\end{equation}
If $\omega$ is small compared with $k_r$ and one takes into account
that the Fourier amplitudes $V_m$ decay rapidly with growth of $m$,
the resonant condition (\ref{res21}) has the simplest form
\begin{equation}
p_\text{max}(I)=\frac{k_r}{k_z}=\frac{\lambda_z}{\lambda_r}.
\label{res22}
\end{equation}
Interplay of vertical and usual ``horizontal'' (i.~e.
relating to the condition (\ref{rescond})) resonances
distorts resonant tores \cite{LichLib}. 
Such a distortion can enforce close resonances to overlap,
that causes emergence of global chaos. 

Under the third regime
behavior of near--axial and steep rays is
qualitatively different.
A perturbation causes strong distortion of the sound-speed profile 
near the channel axis, and the derivative $d\omega/dI$ vanishes in the range
of small values of the action, that is a sufficient condition 
of chaos of near-axial rays \cite{Local}. Furthermore,
fast depth oscillations of a perturbation
lead to occurrence of microchannels near 
the waveguide axis, and instability
is amplified by irregular jumps of rays
from one microchannel to another.

On the other hand, fast oscillations of a perturbation along steep rays can be eliminated 
using the averaging technique and, therefore, the problem can be reduced
to the integrable one \cite{LanLif, Zas, Rahav}.
The averaging technique the
spatial frequency $\omega$ to be small compared with
$|k|$. Certainly, this condition doesn't hold near the refractive ray turning points
where $k=0$ (under reflections from the surface 
$|p|\ne 0$ and $k$ can be large enough).
In these regions influence of horizontal resonance (\ref{rescond}) has to be taken into account.
Nevertheless, the size of the ``resonant'' region, 
being estimated as $\Delta p\sim k_z^{-1}$, is small for large $k_z$. 
Moreover, it can be shown that the impact of horizontal resonance
ceases with increasing $k_z$.
Let us rewrite the range-dependent term in the Hamiltonian in the form
\begin{equation}
H_1=V(I,\vartheta)\exp[ik_zz(I,\vartheta)+ik_rr]+\text{c.~c.}.
\label{h1-e}
\end{equation}
Then we get Fourier-expansion of the function $z(\vartheta)$
\begin{equation}
z(\vartheta)=z_0+\sum\limits_{n=1}^{\infty} z_n\cos{n\vartheta},
\label{q_ryad}
\end{equation}
and substitute it into the equation (\ref{h1-e}). Leaving
the superior term with $n=1$ only, we can represent (\ref{h1-e}) in the form
\begin{equation}
H_1=V_\text{eff}\exp(ik_rr)+\text{c.~c.},
\label{h1-vef}
\end{equation}
where $V_\text{eff}$ is expressed in terms of the Bessel expansion
\begin{equation}
V_\text{eff}=V(I,\vartheta)\exp(ik_zz_0)\sum_{l=-\infty}^\infty i^lJ_l(k_zz_1)\exp(il\vartheta).
\label{bessel}
\end{equation}
According to (\ref{shir}) and (\ref{h1-vef}), the resonant response of a ray can be expressed as
\begin{equation}                    
\Delta I\sim\sqrt{V_\text{eff}}.
\label{resp}
\end{equation}
In the limit $k_zz_1\to\infty$
the Bessel functions have the asymptotics 
\begin{equation}
J_l(k_zz_1)\sim\frac{1}{\sqrt{k_zz_1}}.
\label{asympt}
\end{equation}
Therefore, 
the resonant response $\Delta I$ decreases with increasing $k_z$ as $k_z^{-1/4}$.

Thus the criterion of applicability of the averaging technique
can be formulated in the following way: inequalities
\begin{equation}
\left|\frac{dV}{dz}\right|\ll\left|k_zV\right|,\quad
\omega\ll k_z|p|,
\label{crit}
\end{equation}
must be satisfied for that range of the depth, where the function
$V(z)$ differ considerably from 0.
The first inequality in (\ref{crit}) infers insignificance of resonances along the trajectory.
The second one requires the ray momentum to be large enough.
The averaging technique reduces
the ray equation written as
\begin{equation}
\frac{d^2z}{dr^2}=-\frac{1}{c_0}\left(\frac{d(\Delta c)}{dz}
-\frac{d(\delta c)}{dz}\right).
\label{r-tr}
\end{equation}
into the ``smoothed'' form
\begin{equation}
\frac{d^2z}{dr^2}=-\frac{1}{c_0}\left(\frac{d(\Delta c)}{dz}
-\hat M\left[\frac{d(\delta c)}{dz}\right]\right),
\label{rtr}
\end{equation}
where $\hat M$ is some averaging operator.
In the vicinity of the channel axis the spatial frequency $k$ varies slowly
and the averaged term in (\ref{rtr}) can be expressed as follows \cite{Uleysky}
\begin{equation}
\hat M\left[\frac{d(\delta c)}{dz}\right]=\frac{\varepsilon^2k_z^2V}{2k^2}\frac{dV}{dz}.
\label{r-tr2}
\end{equation}
%

%
%

\subsection{Numerical simulation}

\begin{figure*}[!htb]
\includegraphics[width=0.6\textwidth,clip]{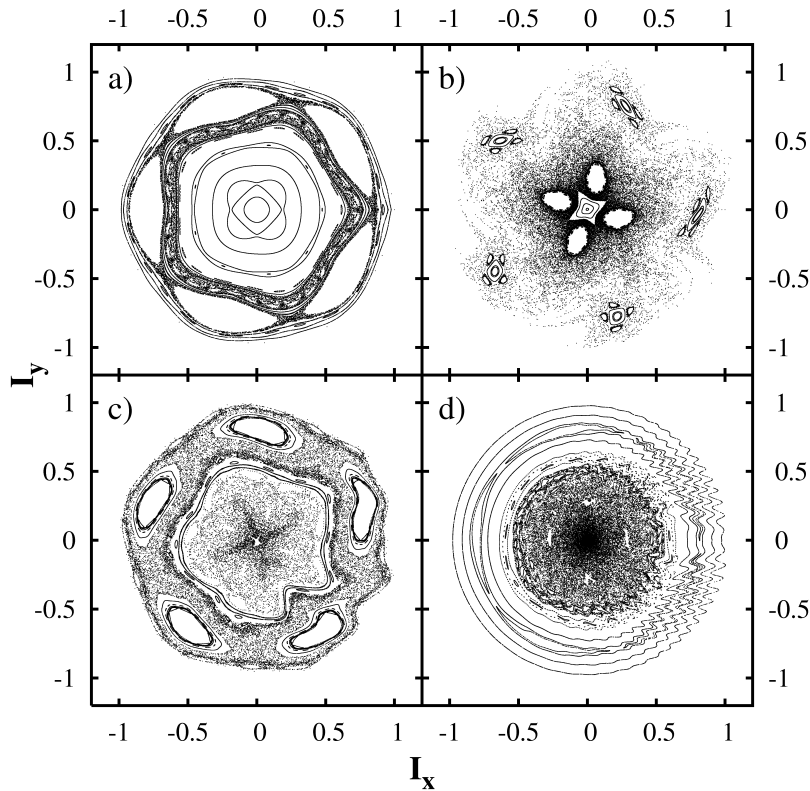}\hfill
\caption{Poincar\'e map in the polar normalized action--angle
variables:
a) $k_z=0$, b) $k_z=2\pi/2$~km$^{-1}$,
c) $k_z=2\pi/1$~km$^{-1}$, d) $k_z=2\pi/0.2$~km$^{-1}$.
$I_x=(I/I_s)\cos{\vartheta}$, $I_y=(I/I_s)\sin{\vartheta}$,
$I_s=I(E=0)$.}
\label{poinc1}
\end{figure*}

The Poincar\'e map is known as an efficient method to separate stable and unstable rays
in the underlying phase space.
We constructed Poincar\'e maps in the polar action-angle coordinates
with different ratios of $k_z$ and $k_r$. 
The result is demonstrated on Fig.~\ref{poinc1}.
$k_z$ varies from 0 to $2\pi/0.2$~km$^{-1}$, while $k_r$ is 
chosen to be of $2\pi/10$~km$^{-1}$ for all computations. 
Other parameters of the perturbation have the following values:
$\varepsilon=0.0025$ and $B=1$~km.

When $k_z$ is very small or equal to zero, ray dynamics is predominantly
regular (Fig.~\ref{poinc1}a).
Chaos occurs only in the isolated chaotic layer
near $I=I(E_r)$, where nondegeneracy violates.
Nevertheless, inclusion of even slow vertical oscillations
broaden the chaotic layer greatly, that is shown in Fig.~\ref{poinc1}b.
The so-called ``chaotic sea'' emerges
and chaotic diffusion in phase space becomes unbounded and global.
The chaotic sea is rarely filled by points due to strong ray escaping. 
Such amplification
of ray escaping is induced by vertical resonance (\ref{res22}) which
affects especially the steep rays satisfying the condition \cite{RAO16}
\begin{equation}
|p_\text{max}(I)|=\frac{\lambda_z}{\lambda_r}=0.2.
\label{vc}
\end{equation}
With further increasing $k_z$ the most steep rays exhibit the transition
chaos -- stability. It is shown in Fig.~\ref{poinc1}c, where the Poincar\'e map 
for the case of $k_z=2\pi/1.0$~km$^{-1}$ is depicted.
Now the chaotic sea is separated into two parts,
the internal layer and the external one (Fig.~\ref{poinc1}c).
Both these layers are restricted by invariant curves.
Complete stabilization of the steep rays takes place
in the case of $k_z=2\pi/0.2$~km$^{-1}$ (Fig.~\ref{poinc1}d).
With that, the internal chaotic layer occupies entirely the range of small values of the action.

\begin{figure}[!ht]
\begin{center}
\includegraphics[width=0.4\textwidth]{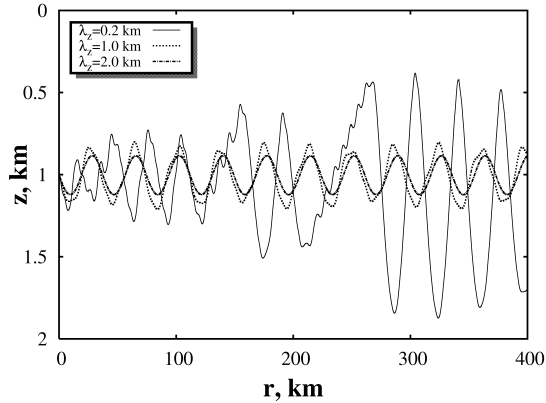}
\caption{
Ray trajectories with different values of vertical 
wavenumber of the perturbation $k_z$. All the rays are launched
from the axis of the waveguide with one and the same
starting momentum $p(r=0)=0.03$ 
}
\label{tr}
\end{center}
\end{figure}

In Fig.~\ref{tr} fragments of three ray trajectories with
$k_z$, varying from $2\pi/2.0$~km$^{-1}$ to $2\pi/0.2$~km$^{-1}$, are shown.
Initial conditions and other parameters of the sound channel are the same for all trajectories.
All the rays are launched from the channel axis with the same
starting momentum $p(r=0)=0.03$.
The ray with $k_z=2\pi/2.0$~km$^{-1}$ is completely stable, that is in agreement with 
Fig.~\ref{poinc1}b.
In the case of $k_z=2\pi/1$~km$^{-1}$ the ray belongs to the internal chaotic layer and,
therefore, is chaotic (this case corresponds to Fig.~\ref{poinc1}c), 
but its irregular behavior is not so apparent, as that 
of the ray with $k_z=2\pi/0.2$~km$^{-1}$.
In the latter case the ``elevating'' effect takes place -- 
the ray, being trapped into a microchannel near the lower turning point,
leaves it only near the upper turning point.
The action of an ``elevated'' ray insreases appreciably.

The qualitative description, being provided by the Poincar\'e maps, enables
to explore the properties of the wave field in the time -- depth plain at a receiver,
the so-called timefront.
Timefronts were the main measurable characteristics in the recent long-range experiments
\cite{Spind}.

Chaotic dynamics of rays leads to divergence of their trajectories and, therefore,
arrival times of chaotic rays with close launching angles can be very different.
Besides that, topology of the Poincar\'e maps (Fig.~\ref{poinc1}) shows that
only the rays with certain values of the action perform chaotic motion,
whereas the rest part of phase space corresponds to stable rays.
Therefore, arrival time spreading is significant only in some finite segments of a timefront
of a received pulse.

\begin{figure}[!ht]
\begin{center}
\includegraphics[width=0.4\textwidth,clip]{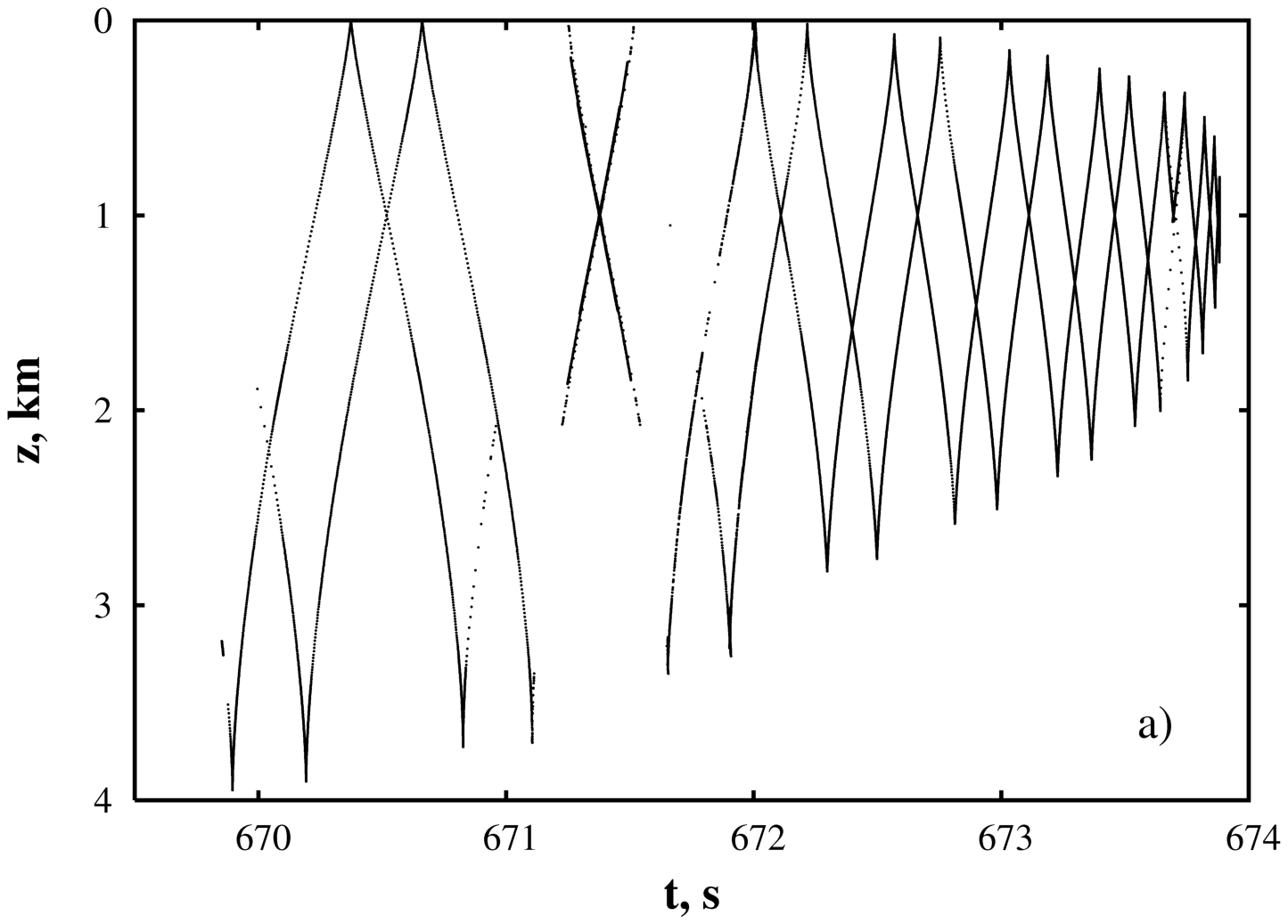}
\includegraphics[width=0.4\textwidth,clip]{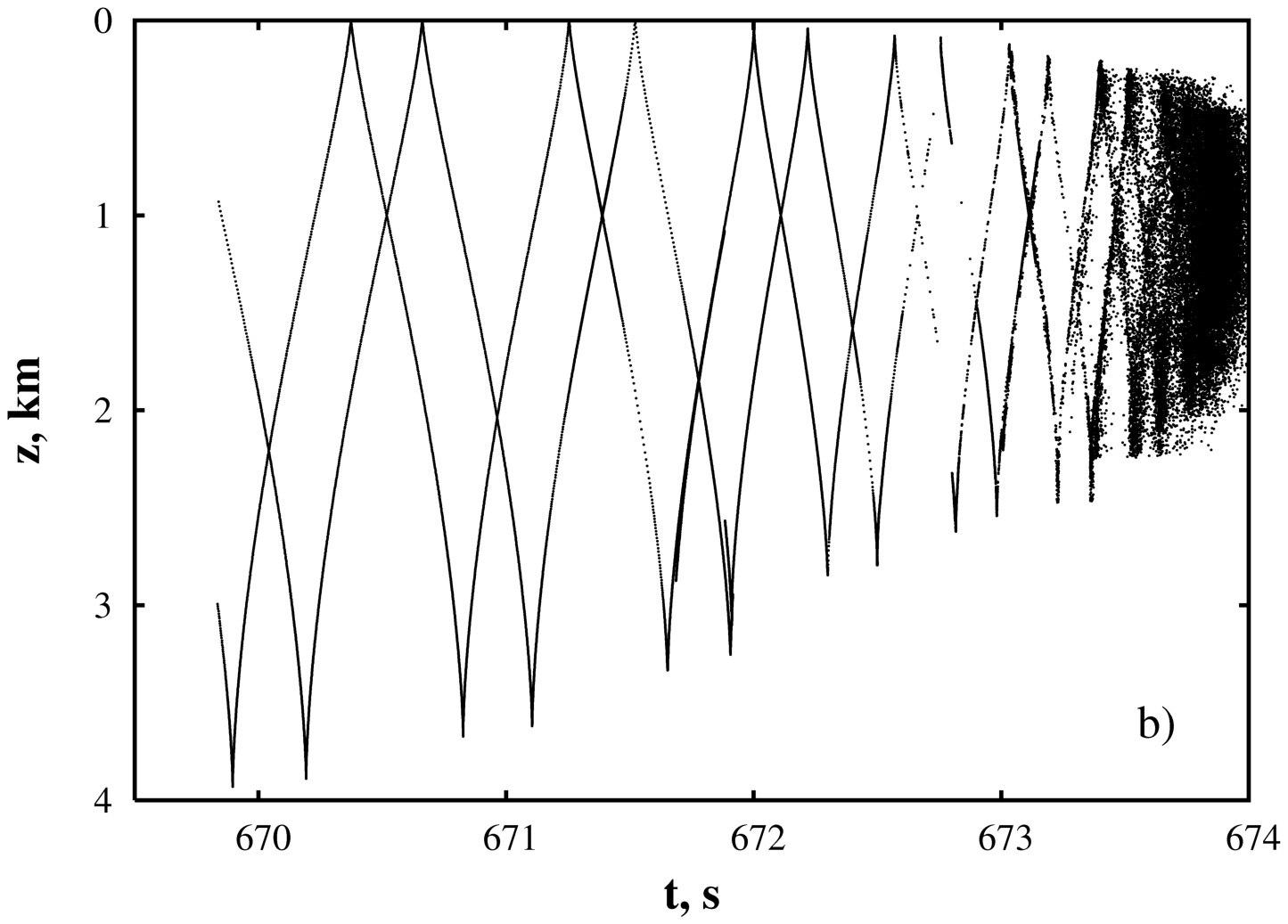}
\caption{
Timefront of a received pulse
at the range 1000~km in the periodically perturbed waveguide:
a) $k_z=0$, b) $k_z=2\pi/0.2$km$^{-1}$.
}
\label{tfrt-per}
\end{center}
\end{figure}

We computed timefronts with different
values of $k_z$. 
Each timefront is modeled as superposition of arrivals of
50000 rays started from the channel axis.
The initial momentum $p$ is uniformly distributed
in the interval $[-p_\text{max}:p_\text{max}]$, where
$p_\text{max}=p(E=0)\simeq 0.22$.
The timefronts for $k_z=0$ and $k_z=2\pi/0.2$km$^{-1}$ are
presented in Fig.~\ref{tfrt-per}.
In the case $k_z=0$ the time spreading is weak. 
There are two prominent gaps and densely filled region amid them in arrival pattern,
those correspond to the horizontal resonance with $l=1$ and $m=5$.
Such gaps are absent in the case of $k_z=2\pi/0.2$km$^{-1}$, that is connected with suppressing of 
horizontal resonance in the regime of fast vertical oscillations of a sound-speed perturbation.
In this case the final portion of the pulse is chaotic and unresolved, that is
a consequence of strong chaos of near-axial rays with small values of the action.

\section{Random sound-speed perturbation}

Internal waves in the deep ocean are known to have continuous spectrum
and must be considered rather as a stochastic process than a deterministic one.
So the following question arises: in what extent results of the previous section
can be referred to acoustic propagation in a randomly perturbed sound channel?
Our goal now is to investigate how ray dynamics depends on the vertical scale
of random longitudinal sound-speed variations.

\begin{figure}[!ht]
\begin{center}
\includegraphics[width=0.4\textwidth,clip]{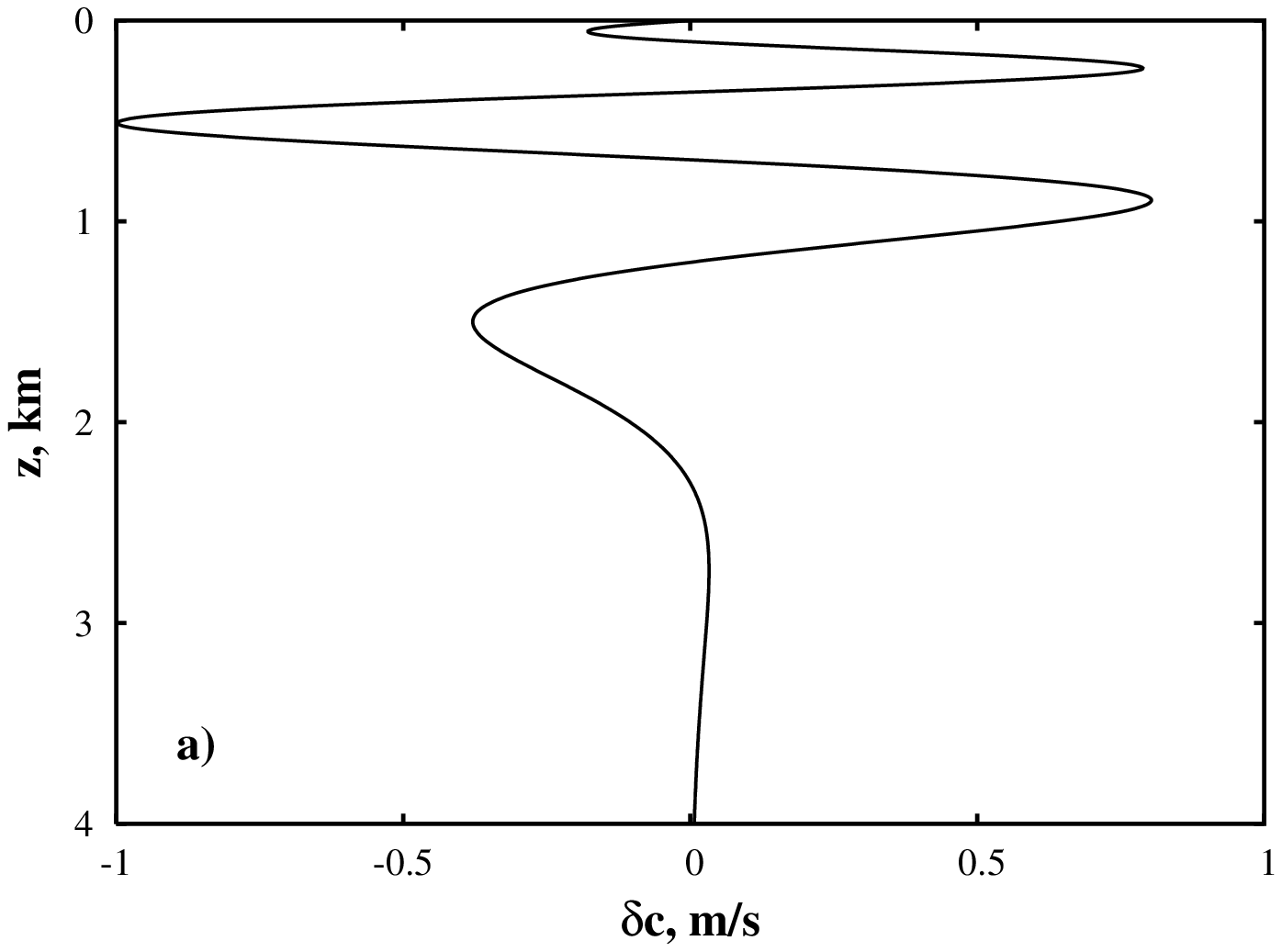}
\includegraphics[width=0.4\textwidth,clip]{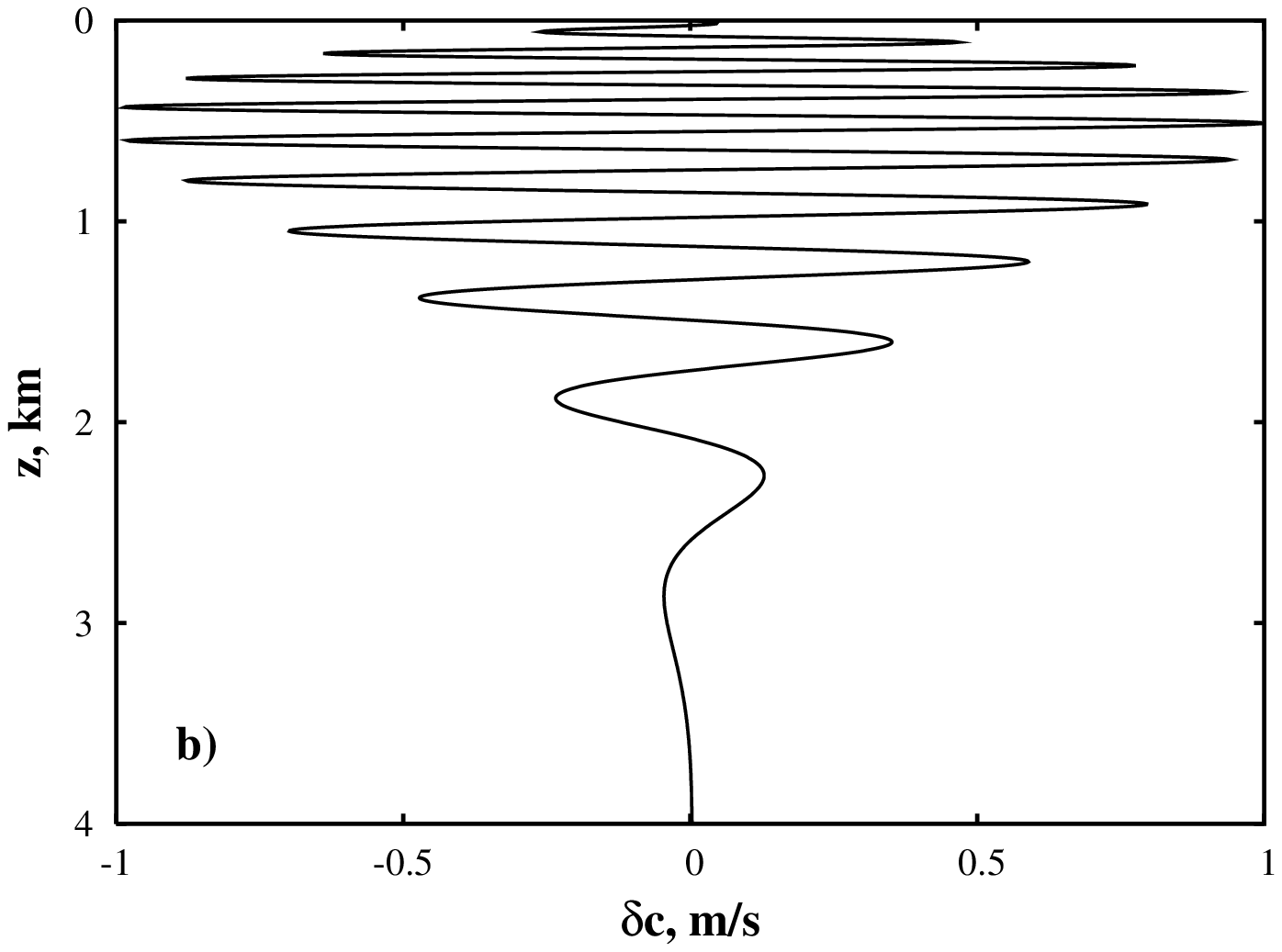}
\caption{
Sound-speed perturbation vs depth:
a) $u=5$, b) $u=20$.}
\label{fig-pert-vmm}
\end{center}
\end{figure}

In the present section we consider the sound-speed perturbation given in the form
\begin{equation}
\delta c(z,r)=\varepsilon c_0\frac{z}{B}\,e^{-2z/B}\xi(z,r),
\label{ver-inh}
\end{equation}
where a random function $\xi(z,\,r)$ is expressed as
\begin{equation}
\xi(z,r)=\cos{\left[\strut\pi\left(ue^{-vz}+
\sqrt{2}A_{\xi}\tilde\xi(r)\right)\right]}\tilde\xi(r),
\label{shum-ver}
\end{equation}
where $A=1$, $v=1$~km$^{-1}$ and
$\tilde\xi(r)$ is a random function with normalized first and second
moments
\begin{equation}
\aver{\tilde\xi(r)}=0,\quad
\aver{\tilde\xi^2(r)}=\frac{1}{2}.
\label{moments}
\end{equation}
A function $\tilde\xi(r)$ is modeled as a sum of 10000 randomly-phased harmonics
with wavenumber distributed in the range from $2\pi/100$~km$^{-1}$ to $2\pi/1$~km$^{-1}$.
Spectral density of $\tilde\xi(r)$ decays with $k$ as $k^{-2}$. 
Single realizations of the perturbation with $u=5$ and $u=20$ are presented in
Fig.~\ref{fig-pert-vmm}.

\begin{figure}[!ht]
\begin{center}
\includegraphics[width=0.4\textwidth,clip]{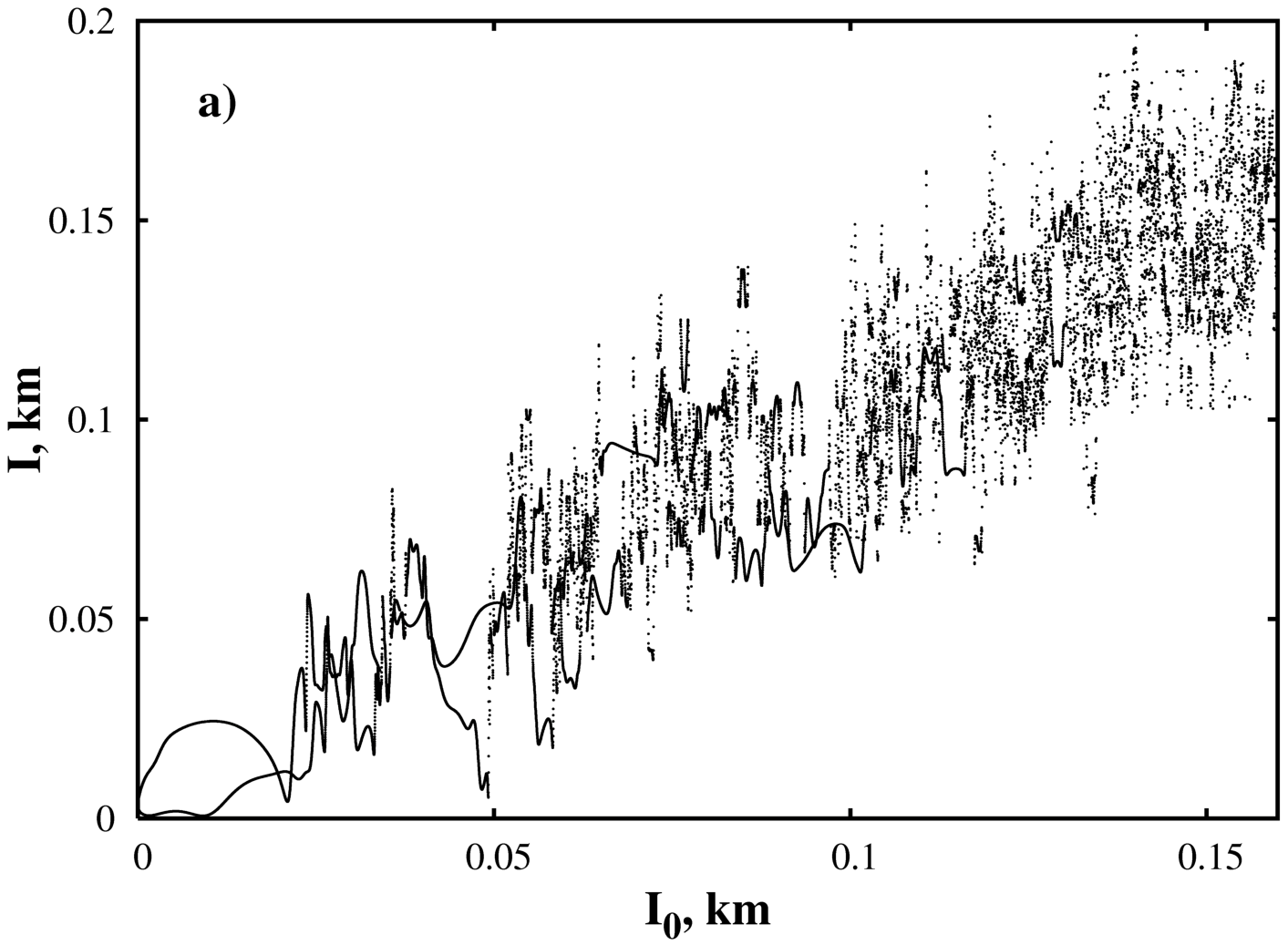}
\includegraphics[width=0.4\textwidth,clip]{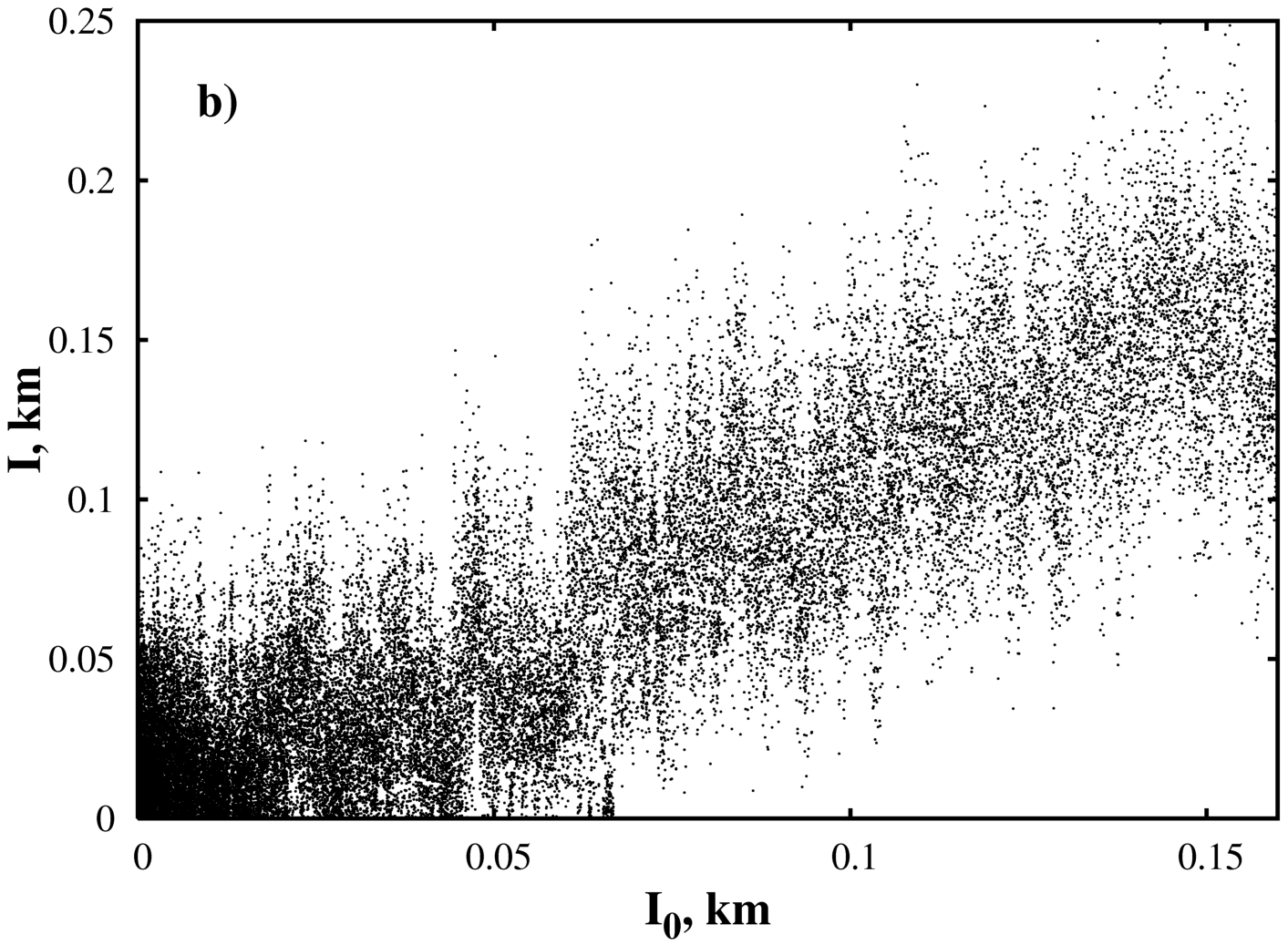}
\caption{
Ray action at the range 1000~km as a function of its initial value
in the randomly perturbed waveguide.
a) $u=5$, b) $u=20$.
}
\label{fig-II0-vmm}
\end{center}
\end{figure}

We integrated numerically the ray equations with the parameter $u$ varying from 5 to 20.
Figure \ref{fig-II0-vmm} represents the ray action at the range
1000~km as the function of the starting action $I_0$.
In the case of slowly varying vertical structure with $u=5$ (upper plot)
chaoticity grows gradually with increasing $I_0$.
Flat rays don't perform extreme sensitivity to initial conditions and
dependence is continuous in the range of small $I_0$.
Isolated continuous pieces belong to bundles formed by
stable or weakly chaotic rays with close launching angles and
converging travel times, namely {\it coherent clusters} \cite{Chaos}.
Coherent clusterization is a sort of cooperative effects
in randomly-driven nonlinear Hamiltonian systems \cite{Arxiv}.
Creation of coherent clusters results from the resonant interaction
of unperturbed ray motion with low-frequency components of a random
perturbation. In point of fact coherent clusters can be considered as counterparts of islands
of stability.
Continuous pieces melt with increasing $I_0$
and the dependence $I(I_0)$ become almost irregular, that attests ray divergence and chaos.

\begin{figure}[!ht]
\begin{center}
\includegraphics[width=0.4\textwidth,clip]{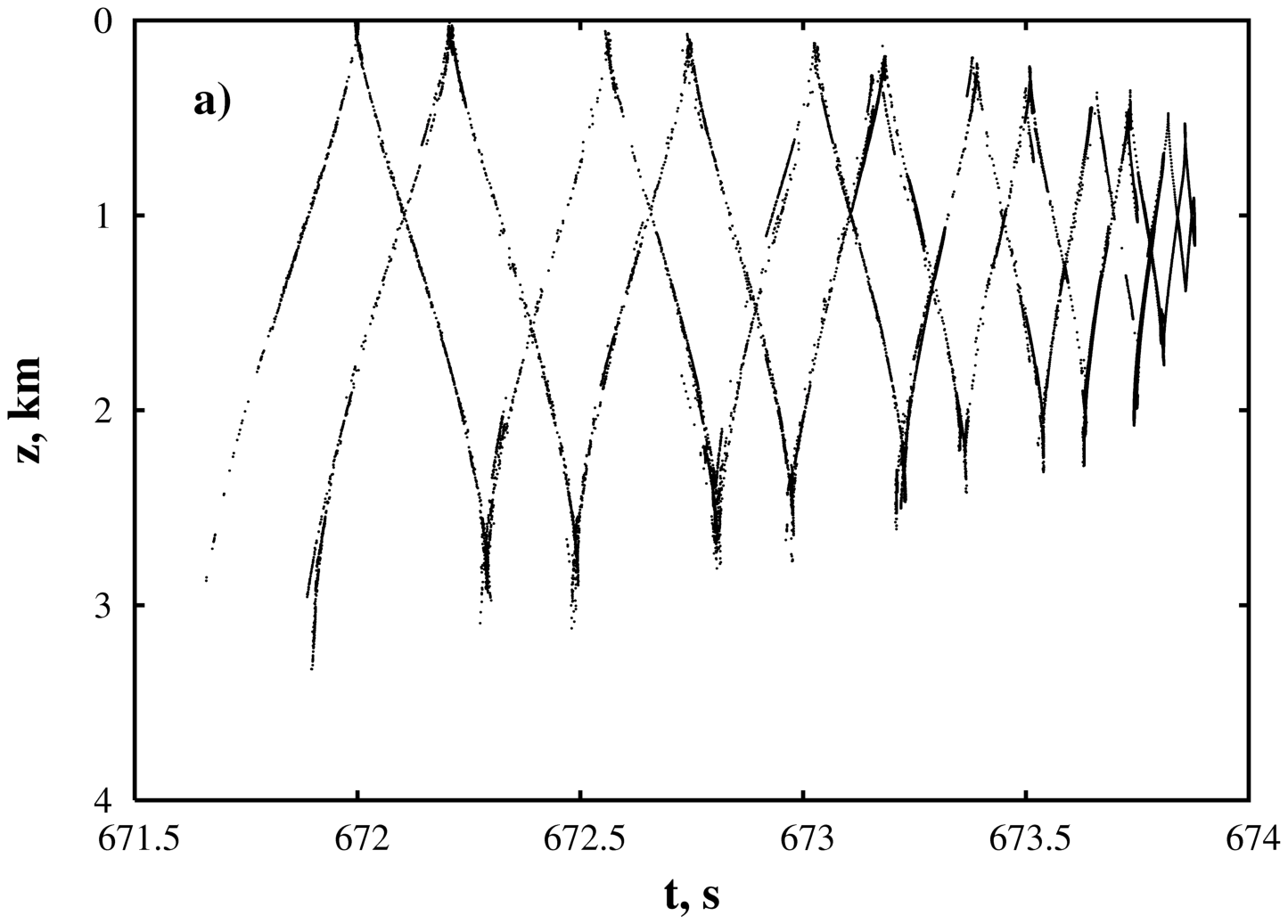}
\includegraphics[width=0.4\textwidth,clip]{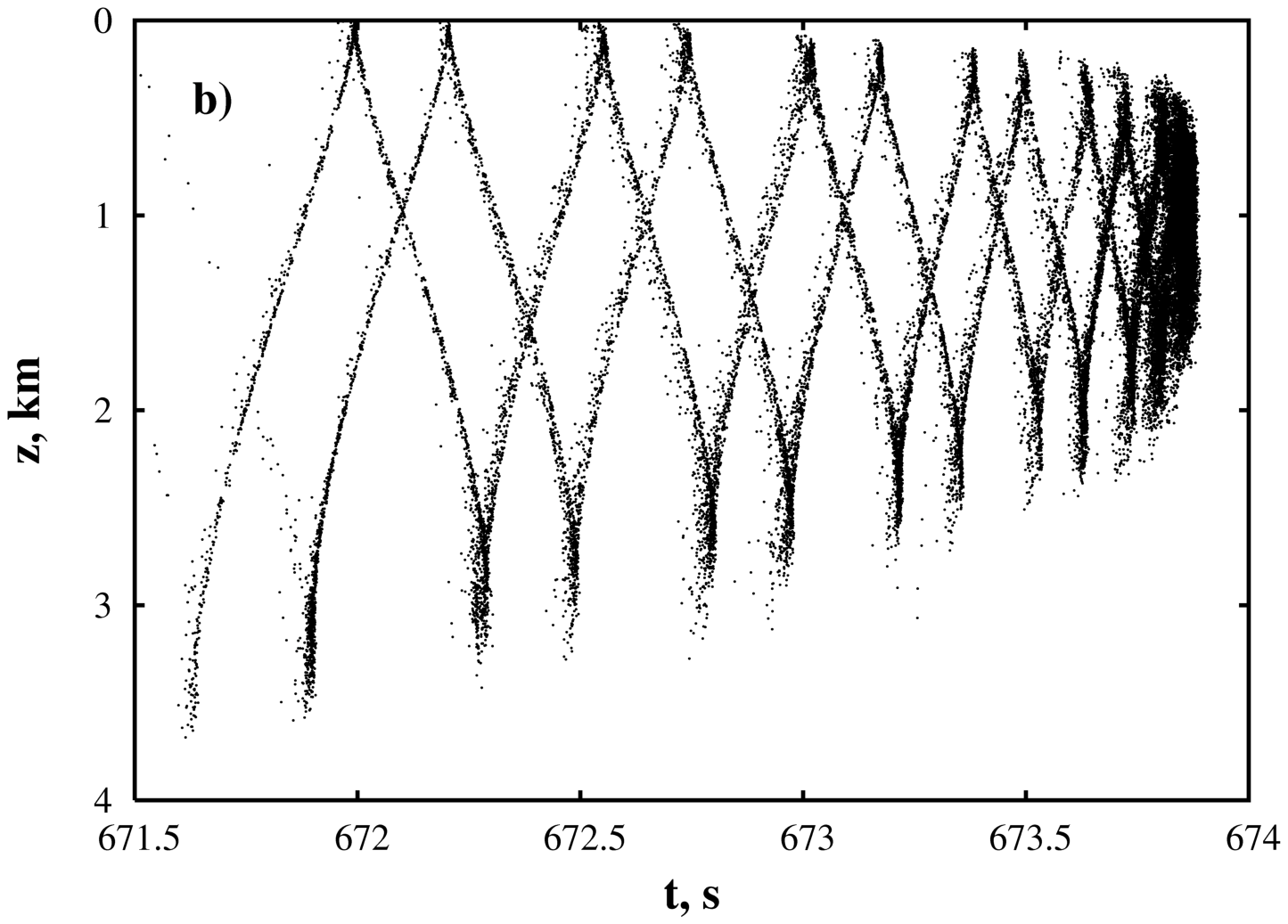}
\caption{
Timefront of a received pulse
at the range 1000~km in the randomly perturbed waveguide:
a) $u=5$, b) $u=20$.
}
\label{fig-tfrt-vmm}
\end{center}
\end{figure}

In spite of the case $u=5$ there are no any remarkable coherent clusters in the case of $u=20$.
The dependence $I(I_0)$ is chaotic for all values of the starting action.
It should be noted that the region of small values of the action
is more fuzzy than the region of large ones.
Moreover, spreading of values of the action  
depends slightly on $I_0$ for steep rays, that implies homogeneous chaotic diffusion.
According to Virovlyansky \cite{Viro, Radio}, if ray diffusion in phase space 
is homogeneous and can be treated as a Wiener process, 
early arrivals are expected to be well-resolved. This statement is confirmed 
by the timefronts computed with $u=5$ and $u=20$ 
(Fig.~\ref{fig-tfrt-vmm}).
If $u=5$ (upper plot)
all the arrivals are well resolved, but their distribution is
nonuniform: coherent clusters, looking as concentrations of plotting points
with extremely small time spreading, alternate with rarely filled regions.
On the other hand,
all the timefront branches are fuzzy in the case of $u=20$.
Time spreading, however, abates with decreasing the arrival time
and is less than distance between neighboring
branches in the early portion of the pulse. 
Locations of branches and the envelope of a timefront in the early portion are coincide
with ones for the unperturbed problem. As long as these characteristics
can be considered as ``marks'' of environment, those are relevant 
for the long-range acoustic thermometry \cite{Worc,Mikh} or 
monitoring of the large-scale ocean variability \cite{RAO15}.

The stability of early portion of a received pulse can be clarified in
the same way as that it has done in the case of the periodic perturbation in the Section III.
Correlation length of sound-speed variations along a ray path 
is given by the formula \cite{BrV}
\begin{equation}
L=\left[\frac{1-p^2}{L_r^2}+\frac{p^2}{L_z^2}
\right]^{-1/2},
\label{correl}
\end{equation}
where $L_r$ and $L_z$ are the horizontal and the vertical correlation 
length scales of sound-speed variations, respectively. If $L_z$ is small enough, 
correlation length is expressed as $L=L_z/p$ and, therefore, is minimal for
steep rays. Thus, if the criterion (\ref{crit}), being expressed as 
\begin{equation}
\left|\frac{dV}{dz}\right|\ll\left|k_zV\right|,\quad
\omega\ll\frac{|p|}{L_z},
\label{crit2}
\end{equation}
is satisfied,
fast oscillations of the perturbation suppress each other and diffusion
in phase space weakens,
that reveals itself in diminishing time spreads in the early portion
of a received pulse.

\section{Summary and discussion}


In the present paper we have investigated in what manner stability of rays
in an underwater sound channel depends on the vertical scale 
of a sound-speed inhomogeneity induced by internal waves.
The main result is that fast depth oscillations
of a sound-speed perturbation can lead to stability of the steep rays 
forming early portion of a received pulse in experiments.
We found that a ray can be captured in resonance with vertical sound-speed variations,
which we call as vertical resonance.
If the length scale of vertical variations is large enough,
vertical resonance enforces strong chaos and escaping of steep rays.

Our results rely on the ray approximation
without referring to the pulse characteristics,
but we suggest that they can be worthwhile for understanding 
some features of the propagation of a monochromatic acoustic wave.
This suggestion is mainly based on the results presented in the recent
paper \cite{Hege}, there the task of the ray/wave method correspondence
was studied. As was reported in \cite{Hege}, improving of
ray predictions requires to smooth those features
of internal wave fine structure, which have vertical length scale 
less than the threshold of wavefield responsibility defined by the formula
\begin{equation}
\lambda_\text{min}=\frac{c_0}{\Omega\tan\phi_\text{max}},
\label{thresh}
\end{equation}
where $\Omega$ is the career frequency and $\tan\phi_\text{max}$ is the maximal
grazing angle. For $\Omega=75$~Hz this scale
of roughly of $110$~m. On the other hand, we have found ray dynamics to be
very sensitive to the fine structure of the internal wave field,
that implies the arrival
pattern to change with varying frequency. For instance,
the early portion of a received pulse can lose stability at very low frequencies,
when the threshold given by (\ref{thresh}) is large.
This seems to associate the anomalous low-frequency
sound attenuation \cite{Vadov} with strong ray escaping.

However, there
are many questions concerning structure of late arrivals.
First, $\lambda_\text{min}$ is large enough for flat rays,
so they are expected to be insensitive to the small vertical features of the
background sound-speed profile.
Second, it is impossible to use geometrical acoustics to describe
the propagation along the axis of a waveguide due to multiple caustics \cite{GFP}.
On the other hand, forming of coherent ray clusters
should explain
the well-resolved coherent late arrivals 
which were observed at 28~Hz in the AST experiment \cite{Wage}.
So the following question arises:
in what extent 
complicated structure of the arrival final 
is linked with ray chaos? 
This topic will be aim of our future research.

\section*{ACKNOWLEDGMENTS}

This work was supported by the project
of Far Eastern Branch of Russian Academy of Sciences ``Acoustical tomography
at long ranges under conditions of ray chaos''.
We wish to thank V.A.~Bulanov, S.V.~Prants and A.O.~Maksimov
for helpful discussion during course of this research.

\section*{Appendix}

The action variable for the rays with $E<E_r$ is given by the formula
\begin{equation}
I=\frac{b}{a}\left(\frac{\mu+\gamma}{2}-\sqrt{\mu\gamma-\frac{2E}{b^2}}
\right).
\label{action1}
\end{equation}
The respective expression for the angle variable is the following
\begin{equation}
\begin{aligned}
\vartheta=\pm
\frac{\pi}{2}\mp\arcsin{\frac{\mu+\gamma-(2\mu\gamma-4E/b^2)\,e^{az}}{Q}},
\end{aligned}
\label{angle}
\end{equation}
where the quantity Q is given by the formula
\begin{equation}
Q=\sqrt{(\mu-\gamma)^2+\frac{8E}{b^2}}.
\label{Q}
\end{equation}
The upper and the lower signs in (\ref{angle}) correspond to the case of $p>0$ 
and to the case $p\le 0$, respectively.
The action and the angle for surface-bounce rays are given by the following formulas
\begin{multline}
I=\frac{p(z=0)}{\pi a}+\frac{b}{a}
\Biggl(
\frac{\mu+\gamma}{4}-\frac{\mu+\gamma}{2\pi}
\times \\ \times
\arcsin{\dfrac{\mu+\gamma-2}{Q}}
-\frac{\pi-\theta_r}{\pi}
\sqrt{\mu\gamma-\frac{2E}{b^2}}
\Biggr),
\label{action2}
\end{multline}
\begin{multline}
\vartheta=
\frac{\pi}{\pi-\vartheta_r}
\Biggl[
\frac{\pi}{2}-\vartheta_r-
\\\shoveright{
-\arcsin{\dfrac{\mu+\gamma-(2\mu\gamma-4E/b^2)\,e^{az}}{Q}}
\Biggr],\quad
p\ge 0,
}
\\
\shoveleft{\vartheta=
\frac{\pi}{\pi-\vartheta_r}
\Biggl[
\frac{\pi}{2}+
}\\+
\arcsin{\dfrac{\mu+\gamma-(2\mu\gamma-4E/b^2)\,e^{az}}{Q}}
\Biggr],\quad
p<0.
\label{angl_ref}
\end{multline}
In (\ref{action2})--(\ref{angl_ref}) we used the notation
\begin{equation}
\vartheta_r=\frac{\pi}{2}-\arcsin\left[\frac{\mu+\gamma-2\mu\gamma+4E/b^2}{Q}\right]
\label{theta_r}.
\end{equation}
Under reflections, the ray momentum
is given by the formula
\begin{equation}
p(z=0)=\sqrt{2E-b^2(\mu-1)(\gamma-1)}.
\label{p0}
\end{equation}

The inverse transformation for the rays, propagating without reflections from the surface,
is expressed as follows
\begin{equation}
z(I,\,\vartheta)=\dfrac{1}{a}\ln
{\dfrac{a^2b^2\,\left(\mu+\gamma-Q\cos{\vartheta}\right)}{2\omega'^2}},
\label{zI}
\end{equation}
\begin{equation}
p(I,\,\vartheta)=
\dfrac{\omega\,Q\sin{\vartheta}}
{a\left(\mu+\gamma-Q\cos{\vartheta}\right)},
\label{pI}
\end{equation}
where 
\begin{equation}
\omega'(I)=\frac{ab\,(\mu+\gamma)}{2} - a^2I.
\label{wI}
\end{equation}
The quantity $\omega'$ coincides
with spatial frequency of ray oscillations $\omega$
for the rays propagating without reflections from the surface.
Position and momentum of the surface-bounce rays are expressed as follows
\begin{equation}
\begin{aligned}
&z(I,\,\vartheta)=\dfrac{1}{a}\ln\dfrac{a^2b^2\left[\mu+\gamma
+Q\cos{\left(\frac{\pi-\vartheta_r}{\pi}(\vartheta+\pi)\right)}\right]}
{2\omega'^2}, \\
&-\pi\le\vartheta\le 0, \\
&z(I,\,\vartheta)=
\dfrac{1}{a}\ln
\dfrac{a^2b^2\left[\mu+\gamma
-Q\cos{\left(\frac{\pi-\vartheta_r}{\pi}\vartheta+\vartheta_r\right)}\right]}
{2\omega'^2}, \\
&0\le\vartheta\le\pi.
\label{zI2}
\end{aligned}
\end{equation}
\begin{equation}
\begin{aligned}
&p(I,\,\vartheta)=
\dfrac{\omega' Q\sin{\left(\frac{\pi-\vartheta_r}{\pi}(\vartheta+\pi)\right)}}
{a\left[\mu+\gamma
+Q\cos{\left(\frac{\pi}{\pi-\vartheta_r}(\vartheta+\pi)\right)}\right]},\\
&-\pi\le\vartheta\le 0, \\
&p(I,\,\vartheta)=
\dfrac{\omega' Q\sin{\left(\frac{\pi-\vartheta_r}{\pi}\vartheta+\vartheta_r\right)}}
{a\left[\mu+\gamma
-Q\cos{\left(\frac{\pi-\vartheta_r}{\pi}\vartheta+\vartheta_r\right)}\right]},\\
&0\le\vartheta\le\pi.
\end{aligned}
\label{pI2}
\end{equation}

\end{document}